\documentclass[reprint,superscriptaddress,amsmath,amssymb,aps,pra]{revtex4-2}
\usepackage{graphicx}
\usepackage{amssymb}
\usepackage{epstopdf}
\usepackage[dvipsnames]{xcolor}
\usepackage[colorlinks=true,
            linkcolor=red,
            urlcolor=blue,
            citecolor=blue]{hyperref}

\newcommand{\RomanNumeralCaps}[1]{\MakeUppercase{\romannumeral #1}}
\begin{document}
 
\title{Engineering impurity Bell states through coupling with a quantum bath}
\author{Tran Duong Anh-Tai}
\email{tai.tran@oist.jp}
\affiliation{Quantum Systems Unit, OIST Graduate University, Onna, Okinawa 904-0495, Japan}
\author{Thomás Fogarty}
\email{thomas.fogarty@oist.jp}
\affiliation{Quantum Systems Unit, OIST Graduate University, Onna, Okinawa 904-0495, Japan}
\author{ Sergi de María-García}  
\affiliation{Instituto Universitario de Matem\'atica Pura y Aplicada, Universitat Polit\`ecnica de Val\`encia, E-46022 Val\`encia, Spain}
\author{Thomas Busch} 
\affiliation{Quantum Systems Unit, OIST Graduate University, Onna, Okinawa 904-0495, Japan}
\author{Miguel A. García-March} 
\email{garciamarch@mat.upv.es}
\affiliation{Instituto Universitario de Matem\'atica Pura y Aplicada, Universitat Polit\`ecnica de Val\`encia, E-46022 Val\`encia, Spain}
\date{\today}   

\begin{abstract}
We theoretically demonstrate the feasibility of creating Bell states in multi-component ultra-cold atomic gases by solely using the ability to control the inter-particle interactions via Feshbach resonances. For this we consider two distinguishable impurities immersed in an atomic background cloud of a few bosons, with the entire system being confined in a one-dimensional harmonic trap. By analyzing the numerically obtained ground states we demonstrate that the two impurities can form spatially entangled bipolaron states due to mediated interactions from the bosonic bath. Our analysis is based on calculating the correlations between the two impurities in a two-mode basis, which is experimentally accessible by measuring the particles positions in the left or right sides of the trap. While interspecies interactions are crucial in order to create the strongly entangled impurity states, it can also inhibit correlations depending on the ordering of the impurities and three-body impurity-bath correlations. We show how these drawbacks can be mitigated by manipulating the properties of the bath, namely its size, mass and intraspecies interactions, allowing to create impurity Bell states over a wide range of impurity-impurity interactions.
\end{abstract}

\maketitle

\section{Introduction}
Great advancements in the development of techniques to experimentally realise, control and measure various quantum systems made from single or few particles are currently fueling a {\it second quantum revolution}~\cite{2018AcinNJP,2020EisertNature,2022FraxanetArxiv}. The resulting quantum technologies rely crucially on the possibility to create and maintain entanglement in these systems~\cite{2009HorodeckiRMP}, and one of the most useful states is the maximally entangled Bell state~\cite{2014BrunnerRMP}. In this state two systems, \textit{A} and \textit{B}, are maximally correlated and each of them holds all possible information about some observable from the other. Therefore creating and controlling correlated quantum states, especially the Bell states, in experimentally accessible systems is of importance and will be beneficial for advancing fundamental science and quantum technologies. 

Cold atom systems have proven over the last two decades that they are very suitable systems to control and study single- and few-particle states~\cite{Cursed1DReview,sowinski2019one}. In particular, the long-standing problem of impurities coupled to an environment has been of immense interest ever since the availability of experimental systems in recent years~\cite{schirotzek2009observation,massignan2014polarons,jorgensen2016observation,hu2016bose,schmidt2018universal,baroni2024mediated,cetina2016ultrafast,cetina2015decoherence,lampo2017bose}. Cold atomic settings allow one to access different parameter regimes and therefore simulate a wider range of interesting phenomena, which are often difficult to study with the same amount of control in condensed matter setups. Impurity physics can be conveniently discussed in terms of polarons, which are quasi-particles that are dressed by the excitations of the surrounding medium~\cite{landau1948effective}. One of the most fascinating aspects of polarons is the induced interactions between them mediated by the quantum many-body environment in which they are immersed (see Ref.~\cite{paredes2024perspective} for a recent review). For example, two uncorrelated impurities which are coupled to the same Bose-Einstein condensate (BEC) can form a bound state (known as a bipolaron) due to the bath-mediated attractive interactions~\cite{2018Camacho-GuardianPRL,2019CharalambousSciPost,2013CasteelsPRA,2018Camacho-GuardianPRX,2021WillPRL,2022JagerNJP}. One of the theoretical frameworks that can successfully describe impurities in ultra-cold gases is the mapping of the BEC-impurity problem to the Fr\"{o}hlich polaron Hamiltonian which describes electron-phonon interactions in condensed matter physics~\cite{tempere2009feynman,kain2014polarons,kain2016generalized,grusdt2016all}. For a more accurate description of impurity physics that takes the exact effects of all interactions into account, one has to go beyond the Fr\"{o}hlich model~\cite{grusdt2017bose}. While in weakly interacting gases the Gross-Pitaevskii theory (see Refs.~\cite{drescher2023medium,schmidt2022self,petkovic2022mediated} and references therein) or the quantum Brownian motion framework~\cite{2019CharalambousSciPost,lampo2017bose,mehboudi2019using,lampo2018non,khan2021quantum} can be used to describe the ground state of the BEC-impurities system, in the strongly interacting regime where particle-particle correlations are prominent, the system has to be solved from first principles. It is also important to note that two recent works have shown that the impurity-bath correlations play a crucial role and result in the suppression of impurity self-localization \cite{breu2024impurities,zschetzsche2024suppression}, which is not captured by the mean-field approach. This demonstrates that a full quantum treatment of the BEC-impurity problem is  necessary, although it is not a trivial task to accurately solve many-body systems with different intra- and inter-species interactions in the continuum due to the exponential growth of Hilbert space. However, in one dimension some exact numerical tools are available and recently the non-classical correlations in stationary and out-of-equilibrium situations have been explored using exact diagonalization methods~\cite{garcia2016entanglement,2018DehkharghaniPRL}, or the multi-layer multi-configuration time-dependent Hartree method for mixtures of identical particles~\cite{2021KeilerPRA,2022TheelPRA,2020MistakidisPRR,2023TheelScipost}. Since the impurities can become entangled with each other due to bath induced interactions, we suggest that it is possible to control correlations between impurities by engineering the interaction with the quantum bath alone. 

\begin{figure}[tb]
\centering
\includegraphics[width=0.8\columnwidth]{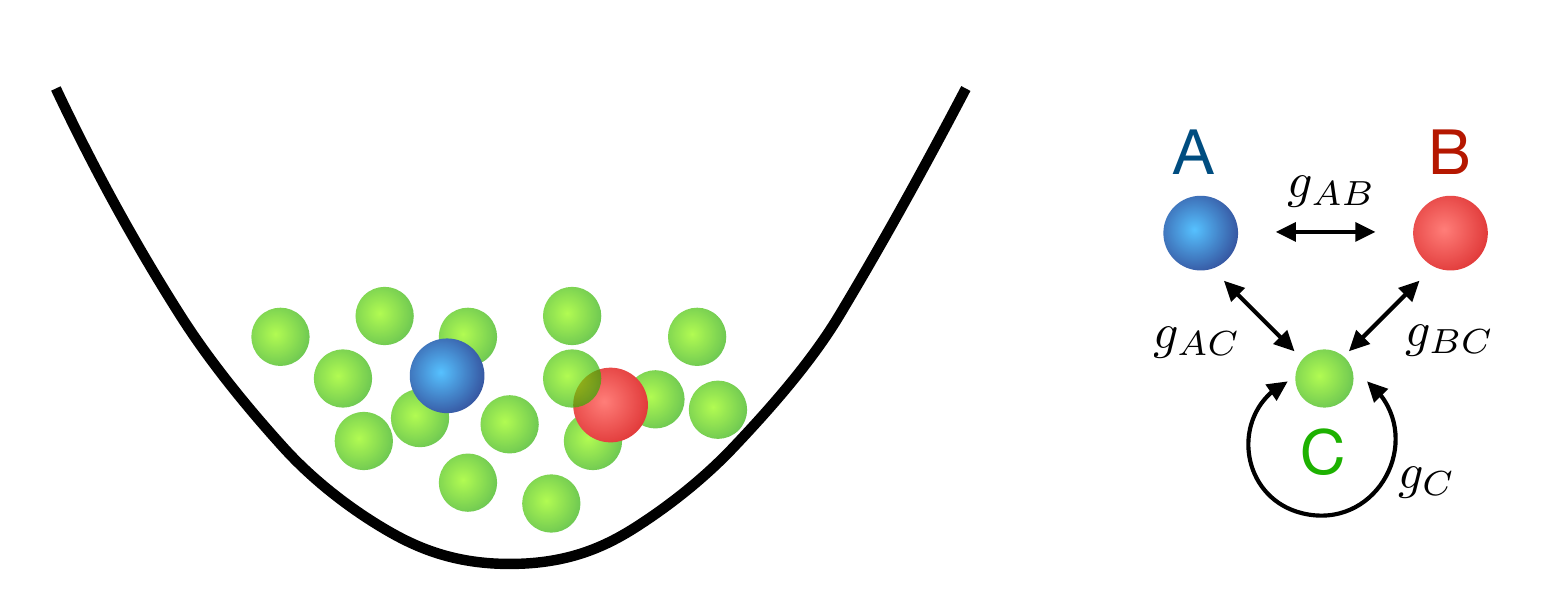}
\caption{Schematic of the model. The $N_{\rm{C}}$ bosons of mass $m_{\rm{C}}$ form an interacting bath that two distinguishable impurities, A and B, of the mass $m$ are immersed in. The entire system is confined in a one-dimensional harmonic trap and all particles interact with all others.}
\label{fig:schematic} 
\end{figure}

In the following we propose a method to engineer strongly-correlated atomic states in ultracold, one-dimensional (1D) interacting atomic systems using the ability to control the inter- and intra-particle interactions via Feshbach resonances. More specifically, we consider two distinguishable impurities immersed into a third larger background component composed of a well-defined number of bosons confined in a harmonic trap (see Fig.~\ref{fig:schematic}). We systematically investigate the ground state of such an interacting three-component mixture by employing the improved Exact Diagonalization scheme~\cite{anhtai2023quantum} to exactly solve the many-body Schr\"{o}dinger equation describing our system (see also Appendix~\ref{app:method}). As a main result of the present paper we demonstrate how to fully engineer non-classical correlations between two distinguishable impurities by tuning the impurity-impurity, and bath intraspecies interactions while the impurity-bath coupling is always repulsive. We find that for strong repulsive interactions between the impurities and the bath, the impurities phase separate to the trap edges, while the bath is localized in the trap center. This allows one to describe the impurities in a discrete spatial basis of the left and right halves of the trap and facilitates the measurement of spatial entanglement and correlation between the impurities~\cite{bergschneider2019experimental}. Our results show that the impurities can form maximally entangled Bell states due to the competition between the different interaction terms, however, the correlations created between the impurities and the bath can impair Bell state formation. This depends on whether the impurities are bunched or anti-bunched, which we explain through the analysis of tripartite correlation functions. Finally, we show how the properties of the bath can be tuned to reduce these higher order correlations, allowing one to create strongly entangled impurity states via bath mediated interactions. 

This paper is structured as follows: Section~\ref{sec:model} presents the model and quantities of interest, while Section~\ref{sec:results_I} discusses how impurity-impurity correlations can be induced by the bosonic bath. In Section~\ref{sec:results_II} we discuss how to significantly enhance the impurity-impurity correlations by engineering the bath and the conclusions and outlook are drawn in Section~\ref{sec:conclusion}.

\section{Model and quantities of interest}
\label{sec:model}
The three-component mixture of interacting ultra-cold bosons we consider consists of two distinguishable atoms of species A and B, which are immersed in a background cloud of $N_{\rm{C}}$ atoms of species C. The atoms A and B are assumed to have the same mass $m_{\rm{A}}=m_{\rm{B}}=m$, while the C atoms have the mass $m_{\rm{C}}$. Since at ultra-cold temperatures the s-wave scattering process is dominant, the interaction between any two particles can be described by a contact interaction potential~\cite{huang1957quantum}, and we assume that the strengths of all intra- and inter-component scattering lengths can be independently adjusted using Feshbach~\cite{chin2010feshbach} or induced-confinement resonances~\cite{haller2010confinement}. For numerical simplicity and experimental relevance, we also assume that all particles are trapped in a one-dimensional parabolic potential with frequency $\omega$, and hence, the Hamiltonian of this system can be written as $\hat{\mathcal{H}} = \hat{\mathcal{H}}_\sigma + \hat{\mathcal{W}}_{\rm{C}} + \hat{\mathcal{W}}_{\sigma\delta}$ where
\begin{align}
\label{full_hamiltonian_first}
    \hat{\mathcal{H}}_\sigma &= \sum_{\sigma} \sum_{i=1}^{N_\sigma} \left[-\dfrac{\hbar^2}{2m_\sigma}\dfrac{d^2}{dx_{\sigma,i}^2} + \dfrac{1}{2}m_\sigma\omega^2 x_{\sigma,i}^2 \right],\nonumber\\ 
    \hat{\mathcal{W}}_{\rm{C}} & = \dfrac{1}{2}g_{\rm{C}} \sum_{i<j}^{N_{\rm{C}}}\delta(x_{{\rm{C}},i}-x_{{\rm{C}},j}),\nonumber\\
    \hat{\mathcal{W}}_{\sigma\delta} & = g_{\sigma\delta} \sum_{i=1}^{N_\sigma}\sum_{j=1}^{N_\delta}\delta(x_{\sigma,i}-x_{\delta,j}).
\end{align}
Since we consider only one atom each of species A and B ($N_{\rm{A}}=N_{\rm{B}}=1$) our setting resembles a two-impurity problem. We therefore term the component C as a bosonic bath, within which the intra-species interactions are characterized by $g_{\rm{C}}$. The inter-species interactions are denoted by $g_{\sigma\delta}$ with $\sigma\neq \delta \in\{\rm{A},\rm{B},\rm{C}\}$ and describe the impurity-impurity interaction $g_{\rm{AB}}$ and the impurity-bath interactions $g_{\rm{AC}}$ and $g_{\rm{BC}}$. In (quasi-)one-dimensional systems, the contact interaction potential can be modeled by a bare $\delta$-function~\cite{olshanii1998atomic}. For the sake of simplicity, we rescale all lengths, energies, and coupling strengths by harmonic oscillator units given by $\sqrt{\hbar/(m\omega)}$, $\hbar \omega$, and $\sqrt{\hbar^3\omega/m}$, respectively.

Three-component ultra-cold quantum gases have been experimentally created \cite{taglieber2008quantum,wu2011strongly} and their properties can vary widely due to the large parameter space the different interaction strengths span. To make the problem tractable, we therefore restrict the interaction strengths between the immersed atoms and the bath to be equal and repulsive, $g_{\rm{AC}}=g_{\rm{BC}}>0$, while the interaction between the impurities, $g_{\rm{AB}}$, and the intra-species interaction $g_{\rm{C}}$ can be either attractive or repulsive. We also fix the maximum number of C-species bosons to be $N_\text{C}=10$ as this is the limit of our numerical simulations. However, as we will show this bath size is more than sufficient to explore the role of bath mediated interactions and correlations, and is in a similar parameter regime as previous studies~\cite{2023TheelScipost}. 

To determine the ground state $|\text{GS}\rangle$ and explore the quantum properties of the three-component system we will solve Eqs.~\eqref{full_hamiltonian_first} numerically exactly using the improved diagonalization method~\cite{anhtai2023quantum}. 
This numerical tool allows us to explicitly access the quantum correlations via the one-body reduced density matrices (OBDM) and the two-body reduced density matrices (TBDM) defined as
\begin{align}
\label{eq:OBDM}
\rho^{(1)}_{\sigma}(x_{\sigma},x_{\sigma}^\prime) &= \langle \hat{\Psi}_{\sigma}^{\dagger}(x) \hat{\Psi}_{\sigma}(x^\prime) \rangle, \\
\label{eq:TBCF}
\rho^{(2)}_{\rm{\sigma\delta}}(x_{\rm{\sigma}},x_{\rm{\delta}},x^\prime_{\rm{\sigma}},x^\prime_{\rm{\delta}})&=
\langle \hat{\Psi}_{\rm{\sigma}}^{\dagger}(x_\sigma) \hat{\Psi}^{\dagger}_{\rm{\delta}}(x_\delta) \hat{\Psi}_{\sigma}(x^\prime_\sigma) \hat{\Psi}_{\delta}(x^\prime_\delta)\rangle,
\end{align}
where $\hat{\Psi}^{(\dagger)}_{\sigma}(x)$ is the bosonic field operator annihilating (creating) a $\sigma$-type boson at the position $x$ and averages are taken with respect to the many-body ground state, $|\text{GS}\rangle$, of the  Hamiltonian~\eqref{full_hamiltonian_first}. 

\begin{figure*}[tb]
\centering \includegraphics[width=0.9\textwidth]{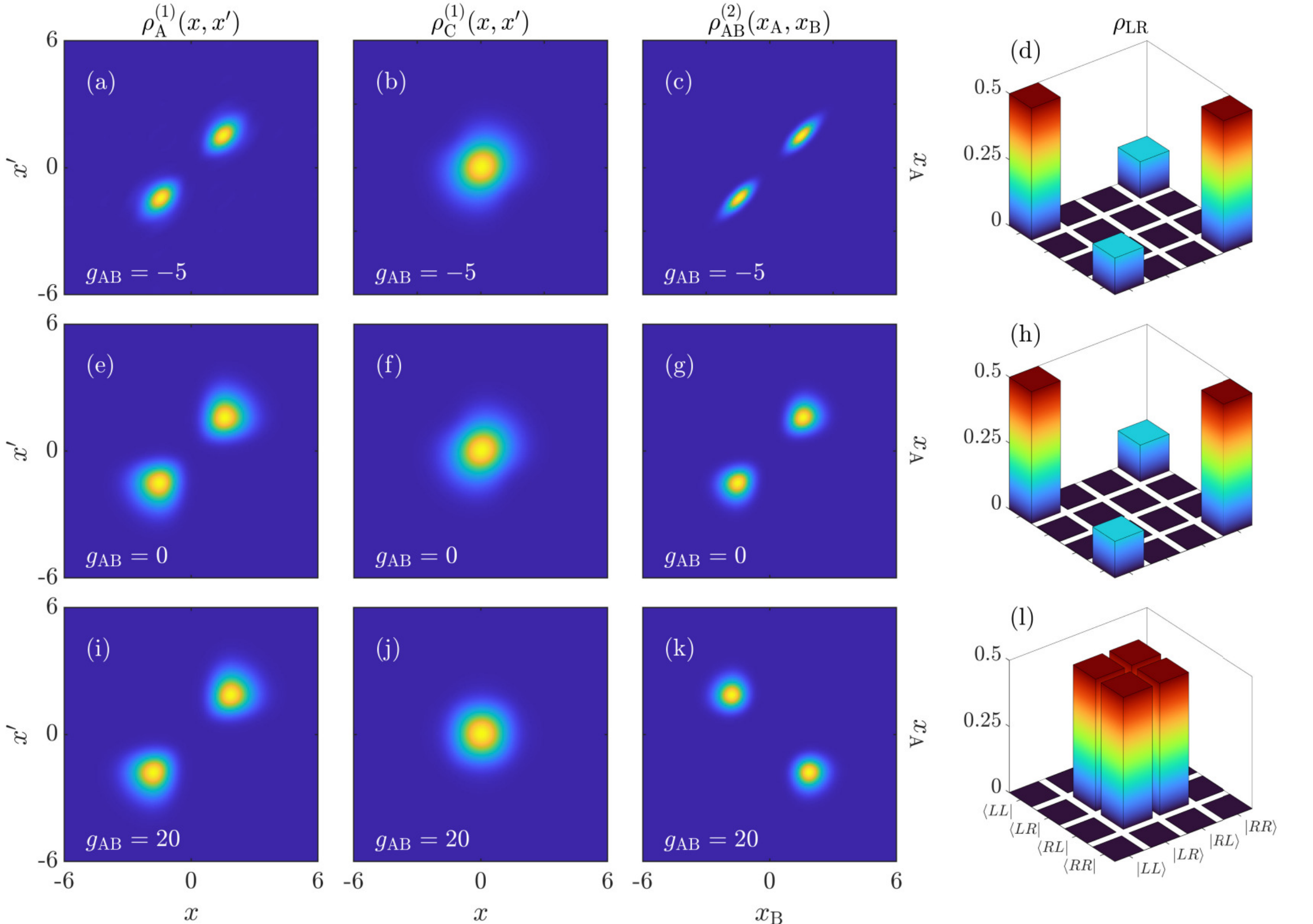}
\caption{The four columns show the OBDM $\rho^{(1)}_{\rm{A}}(x,x^\prime)$, the OBDM $\rho^{(1)}_{\rm{C}}(x,x^\prime)$, the diagonal TBDM $\rho^{(2)}_{\rm{AB}}(x_{\rm{A}},x_{\rm{B}}) = \rho^{(2)}_{\rm{AB}}(x_{\rm{A}},x_{\rm{B}},x_{\rm{A}},x_{\rm{B}})$, and the reduced density matrix $\rho_{\rm{LR}}$ of the many-body ground state of the system for $N_{\rm{C}}=10$. The different rows correspond to different impurity-impurity coupling strengths, $g_{\rm{AB}}=-5$ (upper row), $g_{\rm{AB}}=0$ (middle row), and $g_{\rm{AB}}=20$ (lower row). In all panels, the bosonic bath is non-interacting, $g_{\rm{C}}=0$, while the impurity-bath coupling strengths are kept fixed $g_{\rm{AC}}=g_{\rm{BC}}=5$, and the masses are identical $m_{\rm{C}}/m=1$. Note that all density matrices are normalized to unity.}
\label{fig:figure_2_obtbdm} 
\end{figure*}

\begin{figure}[ht!]
	\centering
	\includegraphics[width=\columnwidth]{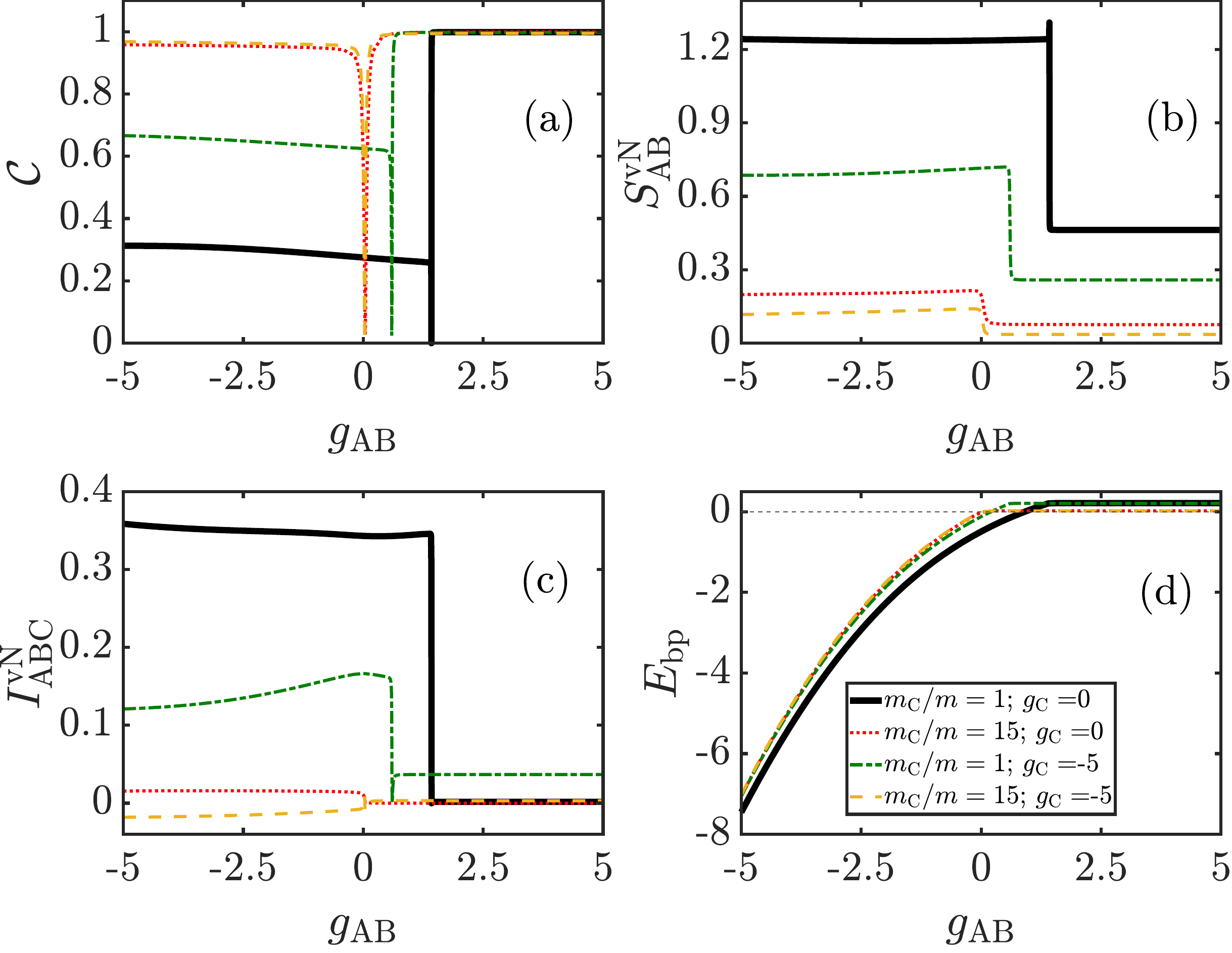}
	\caption{The concurrence (panel a), the von Neumann entropy (panel b), the tripartite mutual information (panel c), and the bipolaron binding energy (panel d) as a function of the impurity-impurity interaction strength $g_{\rm{AB}}$ for $m_{\rm{C}}/m = 1; g_\text{C}=0$ (black line), $m_{\rm{C}}/m = 15; g_\text{C}=0$ (red line), $m_{\rm{C}}/m = 1; g_\text{C}=-5$ (dark-green line), $m_{\rm{C}}/m = 15; g_\text{C}=-5$ (yellow line). The dashed line in panel (d) indicates $E_{\text{bp}}=0$.}
	\label{fig:figure_3_stuffs_vs_gAB}
\end{figure}

\begin{figure}[tb]
\centering
   \includegraphics[width=\columnwidth]{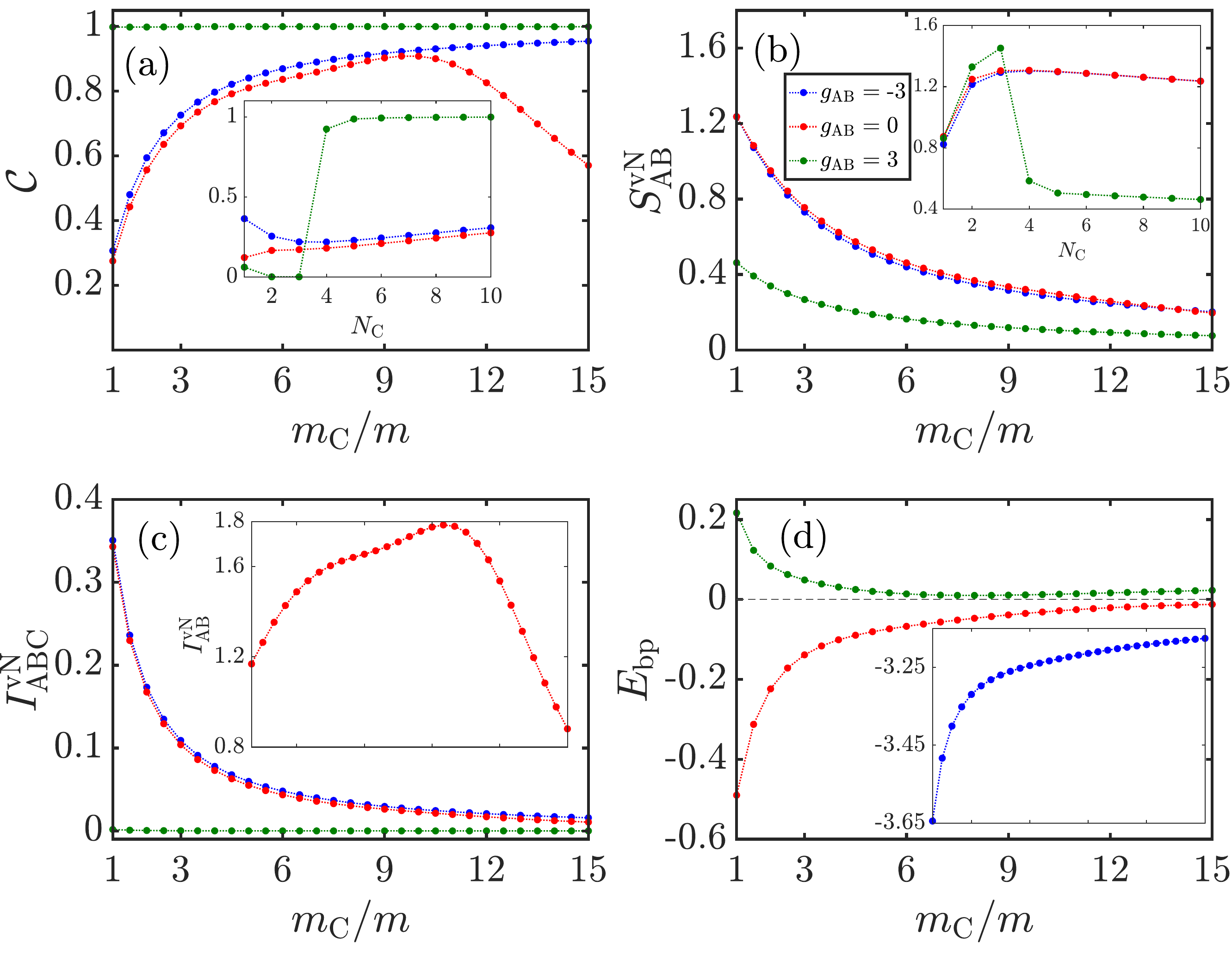}
\caption{Same as Fig.~\ref{fig:figure_3_stuffs_vs_gAB}, but as a function of the mass ratio $m_{\rm{C}}/m$ for the impurity-impurity coupling strength $g_{\rm{AB}} = -3$ (blue line), $g_{\rm{AB}} = 0$ (red line), and $g_{\rm{AB}} = 3$ (dark-green line). The insets in panel (a) and (b) show the concurrence $\mathcal{C}$ and the von Neumann entropy $S_\text{AB}^\text{vN}$ as a function of $N_\text{C}$ for the mass ratio $m_{\rm{C}}/m=1$, respectively. Meanwhile, the inset in panel (c) depicts the bipartite mutual information between two impurities as a function of the mass ratio for $g_{\rm{AB}} = 0$. Note that the horizontal dashed line in panel (d) shows $E_{\text{bp}}=0$ and the x-axis range of insets in panels (c) and (d) is the same as that of the corresponding panels. In all panels, the bath is non-interacting, $g_\text{C}=0$, and has $N_\text{C}=10$ bosons.}
\label{fig:figure_4_stuffs_vs_massratio} 
\end{figure}

To structure our results, we first fix some of the parameters by choosing the impurity-bath coupling as $g_{\rm{AC}}=g_{\rm{BC}}=5$ and considering the masses of all the particles to be identical $m_{\rm{C}}/m=1$.  This allows us to focus on the following three representative cases for the impurity-impurity interaction: case \RomanNumeralCaps{1} where $g_{\rm{AB}}<0$, case \RomanNumeralCaps{2} where $g_{\rm{AB}}=0$, and case \RomanNumeralCaps{3} where $g_{\rm{AB}}>0$. In the two left columns of Fig.~\ref{fig:figure_2_obtbdm} we show the OBDMs of the A (the one for B is identical for symmetry reasons) and the C components. One can see that for all different impurity-impurity interactions the impurities localize towards the trap edges, while the particles of the bath are located in the center of the trap. The total system is therefore in a regime of phase-separation due to the large impurity-bath interactions and the size of the bath. While this describes the spatial separation of the impurities, an analysis of the diagonal of the  TBDM, $\rho_{\rm{AB}}^{(2)}(x_{\rm{A}},x_{\rm{B}})$ (shown in the third column of Fig.~\ref{fig:figure_2_obtbdm}), reveals that the joint probability of finding impurities A and B on either side of C is strongly dependent on the impurity-impurity interaction. Specifically, the impurities anti-bunch for repulsive interactions (case \RomanNumeralCaps{3} $g_{\rm{AB}}=20$) and bunch for attractive interactions (case \RomanNumeralCaps{1} $g_{\rm{AB}}=-5$). Bunching is also apparent for case \RomanNumeralCaps{2}, when the impurity-impurity interaction is absent, as a bound state is formed due to induced attractive impurity-impurity interactions mediated by the C component. This distinctive bound state is known as the bipolaron~\cite{2018Camacho-GuardianPRL,2013CasteelsPRA,2018Camacho-GuardianPRX,2021WillPRL,2022JagerNJP}. The TBDM in Fig.~\ref{fig:figure_2_obtbdm}(g) therefore suggests the combination of bath mediated attractive interactions and suitably large repulsive impurity-bath interactions allows one to create a superposition state of a bipolaron localized at each side of the C component, and that this state has a large degree of spatial entanglement. We will explore this in more detail in the next section.

\section{Bath induced impurity-impurity correlations}
\label{sec:results_I}
To quantify the entanglement between the impurities we first note that in all cases species C can be seen to play the role of a matter-wave barrier located in the center of the trap. This allows us to characterize the impurity states using a discrete spatial basis $|\rm{L}\rangle$ and $|\rm{R}\rangle$ that represents the left ($x<0$) and right ($x>0$) sides of the trap respectively, while we denote the centrally localized C particles by $|\rm{C}\rangle$. 

The precise state which the two impurities are in can then be determined by considering the reduced density matrix of the impurities $\rho_{\rm{LR}}$, which is a positive semi-definite Hermitian operator whose elements are constructed by integrating the TBDM $\rho^{(2)}_{\rm{AB}}(x_{\rm{A}},x_{\rm{B}},x^\prime_{\rm{A}},x^\prime_{\rm{B}})$ over the corresponding spatial regions, i.e. $|\rm{L}\rangle \in (-\infty,0)$ and $|\rm{R}\rangle \in (0,+\infty)$. In Figs.~\ref{fig:figure_2_obtbdm}(d,h,l) we show the reduced density matrices for the three cases we consider. One can immediately note that Fig.~\ref{fig:figure_2_obtbdm}(l) precisely exhibits the density matrix of the Bell state $|\Psi^+\rangle=\left(|\rm{LR}\rangle+|\rm{RL}\rangle\right)/\sqrt{2}$, which is a strong signature that the impurities are maximally entangled in case \RomanNumeralCaps{3}. However, while in cases \RomanNumeralCaps{1} and \RomanNumeralCaps{2} the occupation of the diagonal elements $|\text{LL}\rangle\langle \text{LL}|$ and $|\text{RR}\rangle\langle \text{RR}|$ indicates that the impurities form tight bound states, the coherence terms $|\text{LL}\rangle\langle \text{RR}|$ and  $|\text{RR}\rangle\langle \text{LL}|$ are significantly diminished. This implies that these states are not the maximally entangled $|\Phi^+\rangle=\left(|\rm{LL}\rangle+|\rm{RR}\rangle\right)/\sqrt{2}$ Bell states, and that some decoherence is being caused by the coupling to the C component. Therefore, correlations between the impurities and the matter-wave barrier have a considerable effect on impurity-impurity entanglement, which is substantially different from double-well potentials created by classical optical fields as used in most current ultra-cold atomic setups \cite{bergschneider2019experimental}. We do stress however, that the matter-wave barrier and the aforementioned induced interactions are essential for the creation of the bipolaron Bell state (case \RomanNumeralCaps{2}), as classical fields alone are not sufficient to create this state. 

To quantify the degree of entanglement of the state $\rho_\text{LR}$ we calculate the concurrence  
\begin{equation}
    \label{eq:concurrence}
    \mathcal{C} = \max({0,\sqrt{\lambda_1}-\sqrt{\lambda_2}-\sqrt{\lambda_3}-\sqrt{\lambda_4}})
\end{equation}
where the $\lambda_j$ are the ascending-order eigenvalues of $\rho_{\rm{LR}}(\sigma_y\otimes\sigma_y)\rho_{\rm{LR}}^*(\sigma_y\otimes\sigma_y)$ with $\sigma_y$ being the Pauli $y$ matrix. Furthermore we must consider the separability of the impurities from the C component. This can be quantified by the von Neumann entropy
\begin{equation}
    \label{eq:vonNeuman_entropy}
    S_{\rm{AB}}^{\rm{vN}} = -\text{Tr}(\rho_{\rm{AB}} \log_2(\rho_{\rm{AB}})),
\end{equation}
where $\rho_{\rm{AB}} =\text{Tr}_{\rm{C}} |\text{GS}\rangle\langle\text{GS}|$ is the reduced density matrix of the impurities after tracing out the bath. The condition for the impurities to be separable from the bath is therefore $S_{\rm{AB}}^{\rm{vN}}=0$. 

In Figs.~\ref{fig:figure_3_stuffs_vs_gAB}(a,b) we show the concurrence and the von Neumann entropy as a function of the impurity-impurity interaction $g_{\rm{AB}}$ for $m_\text{C}=m$ (black lines). In the regime of strong repulsive impurity-impurity interactions to which the case \RomanNumeralCaps{3} belongs  the concurrence is maximal, as implied by the populations of $\rho_{\rm{LR}}$ shown in Fig.~\ref{fig:figure_2_obtbdm}(l), and signifies the appearance of the Bell state $|\Psi^+\rangle$~\cite{bonneau2018characterizing,bergschneider2019experimental,usui2020spin}. It is important to note that in this regime the von Neumann entropy is finite, taking values around $S_{\rm{AB}}^{\rm{vN}}\approx 0.4$, indicating that the impurities are not entirely disconnected from the bosonic bath. Since the species C and the impurities are still entangled to some extent, this is another indicator that the bosonic matter-wave barrier which the impurities induce does exhibit quantum effects and cannot be considered equivalent to barriers formed by classical optical fields. 

For decreasing impurity interaction $g_{\rm{AB}}$ there is a sudden transition in the concurrence as the impurities transition from being anti-bunched to being bunched. For mass-balanced systems (black lines in the following graphs) this transition is signaled by a narrow dip in the concurrence to zero in Fig.~\ref{fig:figure_3_stuffs_vs_gAB}(a) and by a maximum in the von Neumann entropy $S^{\rm{vN}}_{\rm{AB}}$ in Fig.~\ref{fig:figure_3_stuffs_vs_gAB}(b), the latter reflecting increased correlations between the impurities and the bath. Both cases \RomanNumeralCaps{1} and \RomanNumeralCaps{2} lie to the left of this transition and take significantly lower values of the concurrence and also increased correlations with the bath. This implies that there is a trade-off between the impurity-bath correlations and the impurity-impurity correlations, usually referred to as entanglement monogamy \cite{coffman2000distributed}. What is unclear is the stark difference between case \RomanNumeralCaps{1} and \RomanNumeralCaps{3}, with the latter possessing maximal concurrence while still having non-zero correlations with component C.  

To understand this dichotomy we first note the differences in the structure of the bunched (cases \RomanNumeralCaps{1} and \RomanNumeralCaps{2}) and anti-bunched (case \RomanNumeralCaps{3}) states. In the latter the impurities are spatially separated and therefore the probability to find both impurities and a bath particle at the same position is negligible. However, in the case of bunching such three-particle coincidences are more likely, therefore leading to the presence of three-body correlations, and we conjecture that these higher order correlations are responsible for the decoherence of the bipolaron Bell-state. To quantify the three-body correlations we compute the tripartite mutual information (TMI)~\cite{seshadri2018tripartite} 
\begin{equation}
    I^\text{vN}_\text{ABC} = I^\text{vN}_\text{AC} + I^\text{vN}_\text{BC} - I^\text{vN}_\text{(AB)C},
\end{equation}
where $I^\text{vN}_{\sigma\delta}$ is the mutual information between one $\sigma$-species particle and one $\delta$-species particle defined as 
\begin{equation}
    I^\text{vN}_{\sigma\delta} = S^\text{vN}_{\sigma} + S^\text{vN}_{\delta} - S^\text{vN}_{\sigma\delta},
\end{equation}
while $I^\text{vN}_\text{(AB)C}$ denotes the mutual information between the two impurities and one C-species boson
\begin{equation}
    I^\text{vN}_\text{(AB)C} = S^\text{vN}_\text{AB} + S^\text{vN}_\text{C} - S^\text{vN}_\text{ABC}\;.
\end{equation}
Here $S^\text{vN}_{\sigma}$ and $S^\text{vN}_{\sigma\delta}$ are the respective single-particle and two-particle von Neumann entropies, while $S^\text{vN}_\text{ABC}$ is the three-particle von Neumann entropy with one particle from each component A, B, and C. The tripartite mutual information is shown in Fig.~\ref{fig:figure_3_stuffs_vs_gAB}(c), and as expected it vanishes when the impurities are anti-bunched, while it takes finite values when the impurities are bunched. This suggests that tripartite impurity-bath correlations are indeed responsible for decohering the $|\Phi^+\rangle$ state. 

Finally, we note that the transition between bunching and anti-bunching occurs at finite values of $g_{\rm{AB}}>0$, as repulsive impurity-impurity interactions are required to counteract the mediated attractive interaction through the C component. The strength of these interactions can be quantified through the bipolaron binding energy \cite{2018Camacho-GuardianPRL}
\begin{equation}
    E_\text{bp} = E_\text{2} - 2E_\text{1} + E_\text{0} ,
\end{equation}
where $E_\text{2}$ is the ground-state energy of the total system with two impurities, $E_\text{1}$ denotes the ground-state energy of the system with one impurity, and $E_\text{0}$ is the energy of the system without any impurities. This is shown in Fig.~\ref{fig:figure_3_stuffs_vs_gAB}(d), where the binding energy takes negative values when the impurities are bound together, notably occurring for zero impurity-impurity interaction at $g_\text{AB}=0$. The transition between bunched and anti-bunched is signalled by the binding energy taking a constant value, occurring at the same value of $g_\text{AB}$ as the major changes in the concurrence, the von Neumann entropy and the mutual information. We note, however, that in the anti-bunched regime the binding energy does not vanish but instead takes a small but constant positive value, which is due to the finite size of the bath.

\begin{figure}[tb]
\centering
\includegraphics[width=\columnwidth]{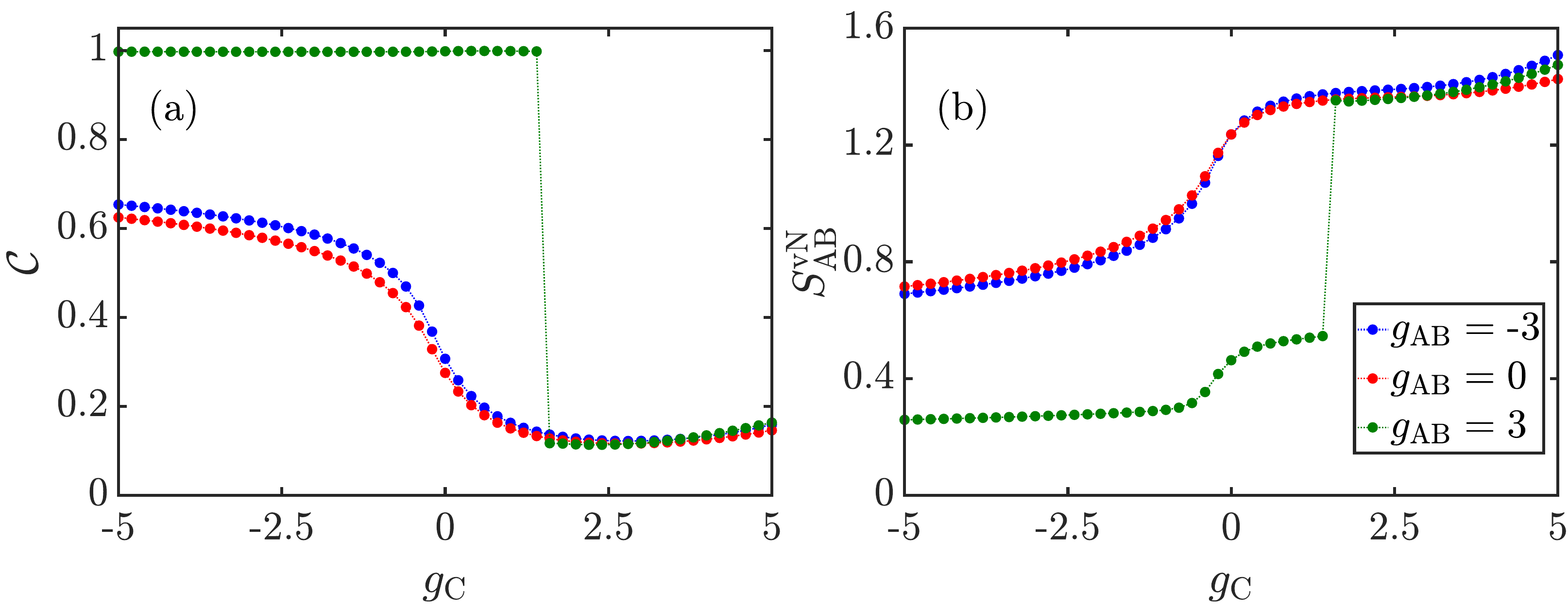}
\caption{The concurrence (panel a) and the von Neumann entropy (panel b) as a function of the intra-component interaction strength $g_{\rm{C}}$ for $g_\text{AB}=-3$ (blue line), $g_\text{AB}=0$ (red line), and $g_\text{AB}=3$ (dark-green line). In all panels the mass ratio is $m_{\rm{C}}/m = 1$.}
\label{fig:figure_5_stuffs_vs_gC} 
\end{figure}

\begin{figure}[tb]
\centering
\includegraphics[width=\columnwidth]{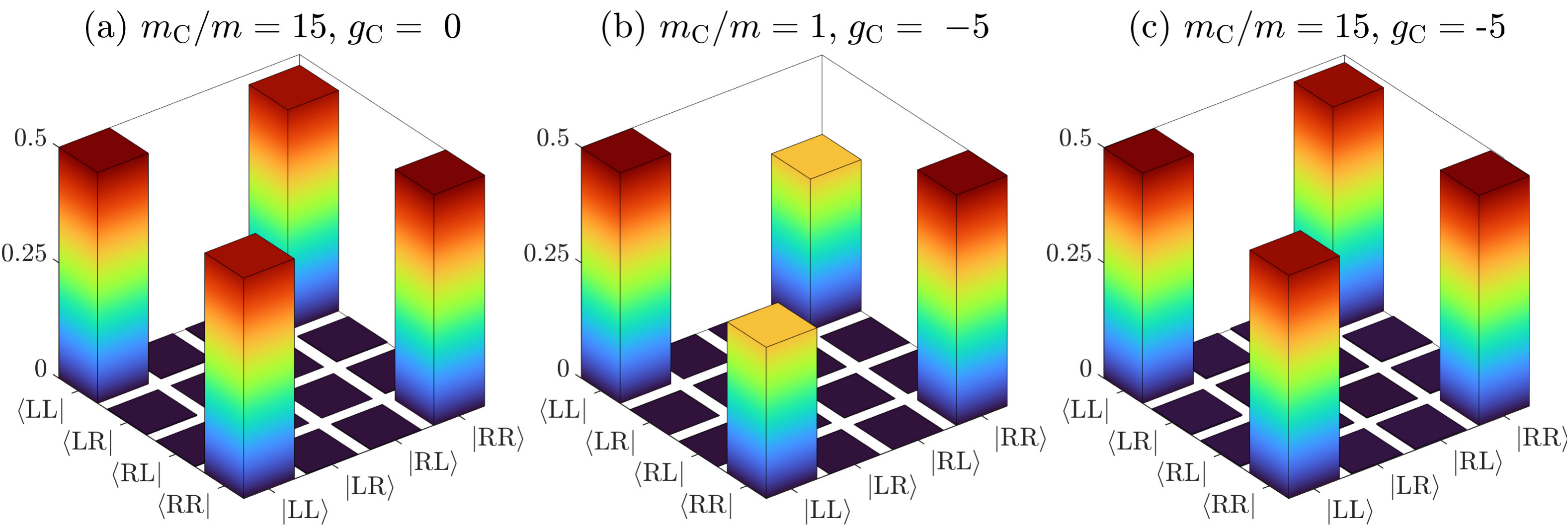}
\caption{The reduced density matrix $\rho_\text{LR}$ for (a) $m_{\rm{C}}/m = 15, g_\text{C}=0$; (b) $m_{\rm{C}}/m = 1, g_\text{C}=-5$; and $m_{\rm{C}}/m = 15, g_\text{C}=-5$. In all panels, the impurity-impurity coupling strength is $g_\text{AB}=-3$, and the bath has $N_\text{C}=10$ bosons.}
\label{fig:figure_6_denmatLR_gC_massratio} 
\end{figure}

\section{Improving impurity-impurity correlations through environment engineering}
\label{sec:results_II}
As we have shown, the coupling to the C component plays a crucial role in spatially separating the impurities from the bath and being able to create the impurity Bell states. On the other hand, the impurity-bath correlations that are created by this coupling also negatively affect the purity of the entangled impurity states and consequently cannot be completely separated from the bath. However, the properties of the bath are also tunable; for instance, by choosing a different atomic species the mass ratio $m_\text{C}/m$ can be varied, while there is also some degree of control over the intraspecies interactions $g_{\rm{C}}$. Let us first focus on adjusting the mass ratio while keeping $g_{\rm{C}}=0$ fixed. The von Neumann entropy, concurrence, mutual information and binding energy are shown in  Fig.~\ref{fig:figure_4_stuffs_vs_massratio} as a function of the mass ratio for fixed impurity-impurity interactions $g_\mathrm{AB}=-3,0,3$. 

It is immediately apparent from looking at the decay of the von Neumann entropy and tripartite mutual information that increasing the mass of the bath particles can significantly reduce the degree of the impurity-bath correlations. This can be well described by an algebraic decay of the form $S^\text{vN}_\text{AB}\sim \left(m_\text{C}/m\right)^{-\alpha}$ and we show the fitted decay rates in Table \ref{tab:DecayRates}. It is worth mentioning that the decrease of the bath-mediated attractive interactions due to the larger mass of the bath particles compared to the impurities is in good agreement with the prediction in Ref.~\cite{2018Camacho-GuardianPRX}. The same effect, albeit with a smaller decay rate, can be achieved by increasing the number of particles $N_\text{C}$, while keeping $m_\text{C}/m=1$ (see inset of Fig.~\ref{fig:figure_4_stuffs_vs_massratio}(b)), with the fits to $S^\text{vN}_\text{AB}\sim N_\text{C}^{-\beta}$ shown again in Table \ref{tab:DecayRates}. It is evident that the exponent $\alpha$ is almost unaffected by the interaction strength between the two impurities, while the decay rate $\beta$ can be increased in the regime of repulsive impurity-impurity interactions. We also note that there is a minimum number of \text{C} particles that are needed to spatially separate the impurities, for instance $N_\text{C}>3$ for $g_\text{AB}=3$, as for smaller bath sizes the impurities are bunched in the center of the trap and hence the concurrence is roughly zero and the von Neumann entropy takes significantly larger values as shown in the insets in panel (a) and (b) in Fig.~\ref{fig:figure_4_stuffs_vs_massratio}.

The enhancement of the correlations for the bunched states in case \RomanNumeralCaps{1} can therefore be attributed to the screening of the bath-mediated interactions by the heavier mass of the bath which in turn reduces impurity-bath correlations and increases the coherence terms in $\rho_\text{LR}$, see Fig.~\ref{fig:figure_6_denmatLR_gC_massratio}(a). This screening effect may be quantified by a reduction in the bipolaron binding energy as shown in Fig.~\ref{fig:figure_4_stuffs_vs_massratio}(d), and can be nicely understood by considering the special case of the bipolaron ($g_\text{AB}=0$). In this case the impurities can only be correlated through the bath, therefore any impurity-impurity correlations are a by-product of impurity-bath correlations. The decrease of the concurrence when $m_\text{C}>10m$ is then a consequence of the reduction of the bath mediated interactions which are insufficient to tightly bind the impurities, leading to finite populations in $|\text{LR}\rangle$ and $|\text{RL}\rangle$. This decay can be further quantified by the bipartite mutual information $I_\text{AB} = S^\text{vN}_\text{A} + S^\text{vN}_\text{B} - S^\text{vN}_\text{AB}$ \cite{bellomia2024quasilocal,amico2008entanglement} (see inset of Fig.~\ref{fig:figure_4_stuffs_vs_massratio}(a)), which describes the continuous variable entanglement between the impurities. The mutual information similarly has to vanish when the impurities and bath become separable, and is therefore a more intuitive measure of the strength of the mediated attractive interactions in the system. We also highlight details of the screening effect of the increased mass ratio $m_\text{C}/m=15$ in Fig.~\ref{fig:figure_3_stuffs_vs_gAB} (red dotted lines) as a function of $g_\text{AB}$. One can see that, when compared to the mass balanced case (black lines), the bunching to anti-bunching transition is shifted towards $g_\mathrm{AB}=0$ as the bath mediated interactions are significantly diminished.

\begin{table}
\caption{\label{tab:DecayRates} Fitted exponents $\alpha$ and $\beta$. The number of bath particles used for fitting $\alpha$ is $N_\mathrm{C}=10$, while the mass ratio for fitting $\beta$ is fixed at $m_\mathrm{C}/m = 1$. To exclude any finite-size effects, we only consider values of $m_\mathrm{C}/m \geq 2$ when determining $\alpha$. Similarly, we use values of $ 6 \leq N_\mathrm{C} \leq 10$ when determining $\beta$. In both cases the fitting equation is of the form $y = ax^{-r}$. }

\begin{ruledtabular}
\begin{tabular}{p{1cm}|p{2cm}|p{2cm}|p{2cm}}
                       &  $g_\mathrm{AB} = - 3$         &        $g_\mathrm{AB} = 0$  &     $g_\mathrm{AB} = 3$    \\ 
\hline
$\alpha$   &   0.74                      &        0.73              &     0.72                \\ 
\hline
$\beta$     &   0.077                     &        0.079             &    0.13               \\ 
\end{tabular}
\end{ruledtabular}
\end{table}

\begin{figure*}[tb]
    \centering
    \includegraphics[width=\linewidth]{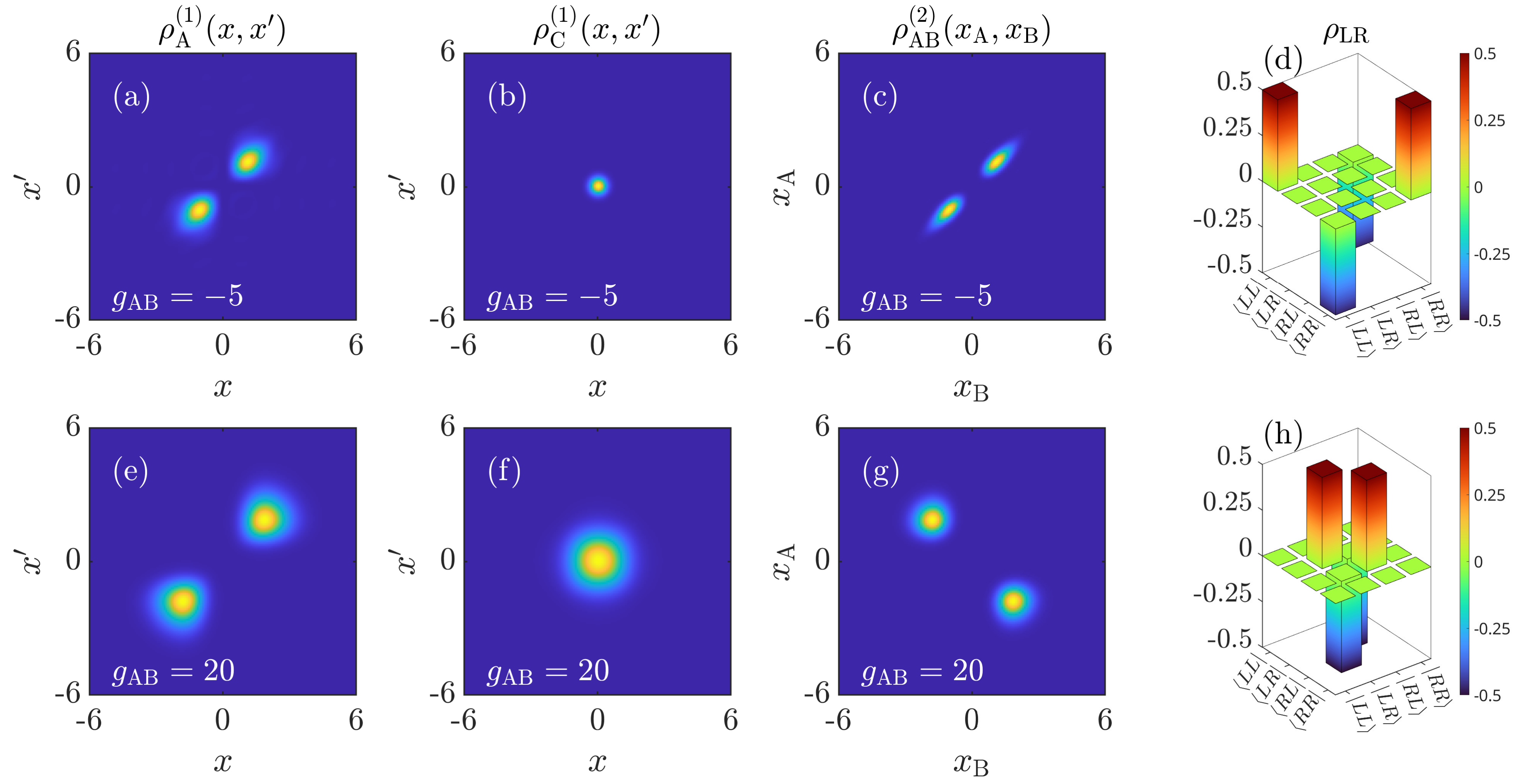}
    \caption{The Bell states $|\Psi^-\rangle$ and $|\Phi^-\rangle$ are the first excited state of the corresponding system whose ground state demonstrates the states $|\Psi^+\rangle$ and $|\Phi^+\rangle$. The four columns are the OBDM $\rho^{(1)}_{\rm{A}}(x,x^\prime)$, the OBDM $\rho^{(1)}_{\rm{C}}(x,x^\prime)$, the diagonal TBDM $\rho^{(2)}_{\rm{AB}}(x_{\rm{A}},x_{\rm{B}}) = \rho^{(2)}_{\rm{AB}}(x_{\rm{A}},x_{\rm{B}},x_{\rm{A}},x_{\rm{B}})$, and the reduced density matrix $\rho_{\rm{LR}}$, respectively. The different rows correspond to different cases, $g_{\rm{AB}}=-5$, $m_{\rm{C}}/m=15$ (upper row), and $g_{\rm{AB}}=20$, $m_{\rm{C}}/m=1$ (lower row). In all panels, the bosonic bath has $N_{\rm{C}}=10$ non-interacting ($g_{\rm{C}}=0$) bosons, while the impurity-bath coupling strengths are kept fixed $g_{\rm{AC}}=g_{\rm{BC}}=5$. Note that all density matrices are normalized to unity.}
    \label{fig:enter-label}
\end{figure*}

Finally, in Fig.~\ref{fig:figure_5_stuffs_vs_gC} we examine the effect of the bath intraspecies coupling strength, $g_{\rm{C}}$, on the concurrence and von Neumann entropy for fixed $m_{\rm{C}}=m$. Increasing the repulsive interactions between the C component atoms broadens the density distribution of the bath, and leads to an increase in the impurity-bath correlations due to larger wave-function overlap in the finite trap environment. This also adversely affects the concurrence for all the three cases we consider, with the sudden decrease for the case $g_\text{AB}=3$ again a signal of this state transitioning from being anti-bunched to bunched. On the other hand, for attractive interactions $g_{\rm{C}}<0$ the concurrence is increased and the von Neumann entropy reduced, which indicates better conditions for the formation of the $|\Phi^{+}\rangle$ Bell state. Since the attractively interacting environment plays a role similar to the one of increasing mass, albeit to a lesser extent (see green dash-dotted lines in Fig.~\ref{fig:figure_3_stuffs_vs_gAB} and $\rho_\text{LR}$ in Fig.~\ref{fig:figure_6_denmatLR_gC_massratio}(b)), a combination of both effects can be used to enhance the impurity-impurity correlations. To highlight this we show in Fig.~\ref{fig:figure_3_stuffs_vs_gAB} how bath interactions $g_\text{C}=-5$ in combination with a mass ratio of $m_\text{C}=15m$ (yellow dashed lines), can be used to significantly improve the impurity-impurity correlations over the whole range of $g_\text{AB}$. Specifically, for $g_\text{AB}=-5$ the concurrence reaches $\mathcal{C}\approx 0.964$ while the von Neumann entropy is $S^{\rm{vN}}_{\rm{AB}}\approx 0.125$. This characterizes a state that is close to the $|\Phi^{+}\rangle$ Bell state and we show $\rho_\text{LR}$ for this case in Fig.~\ref{fig:figure_6_denmatLR_gC_massratio}(c). We also note that the TMI is negative for $g_\text{AB}<0$, indicating that the composite system AB has more information compared to the individual systems A and B \cite{seshadri2018tripartite} due to it forming a tightly bound state. 

It is also worth noting that the spatially anti-symmetric Bell states, $|\Psi^-\rangle=\left(|\rm{LR}\rangle-|\rm{RL}\rangle\right)/\sqrt{2}$ and $|\Phi^-\rangle=\left(|\rm{LL}\rangle-|\rm{RR}\rangle\right)/\sqrt{2}$, naturally appear as the first excited states of the system. Although the OBDM and the diagonal TBDM of the first excited state are quite similar to those of the ground state due to the fact that they are nearly degenerate, we show in Fig.~\ref{fig:enter-label} that a clear difference appears in the reduced density matrix in the $\{|\text{L}\rangle,|\text{R}\rangle \}$ basis $\rho_\text{LR}$ as the coherence terms are necessarily negative. Furthermore, we note that like the spatially symmetric Bell states $|\Psi^+\rangle$ and $|\Phi^+\rangle$, we can also improve the fidelity of $|\Psi^-\rangle$ and $|\Phi^-\rangle$ by modifying the bath parameters such as its mass, giving us qualitatively similar results as discussed previously.

\section{Conclusions}
\label{sec:conclusion} To summarize, we have proposed a scheme for robust preparation of strongly-correlated states with current ultra-cold atomic setups, where particle numbers and interaction strengths are experimentally controllable, and where the spatial correlations we describe can be easily measured. We numerically demonstrate that strongly correlated states close to Bell states can be formed as ground states in systems of two distinguishable impurities immersed in a bosonic reservoir with a well-defined particle number. Our analysis shows that the Bell state $|\Psi^+\rangle$ can be readily generated with two repulsively interacting impurities, while the Bell state $|\Phi^+\rangle$ is hard to achieve due to large correlations created between the bath and impurities which reduce the states coherence. With the aim of reducing the impact of impurity-bath correlations such that the entanglement between the two impurities can be further enhanced, we have investigated the properties of the bath including its intra-species coupling strength, and the mass of its particles. We have demonstrated that in both situations the formation of the Bell states is substantially enhanced compared to the mass-balanced non-interacting bath. Importantly, we have shown in this paper that the bosonic bath located in the center of the harmonic trap forms a matter-barrier to separate the two impurities resulting in the emergence of their non-classical properties. We emphasize that this effect is a purely quantum phenomenon proven by the finite von Neumann entropy between the bosonic bath and impurities and that it is significantly different from classical potentials. We anticipate that our results not only pave an efficient and experimentally feasible way to create and fully control strongly correlated atomic states, but also will stimulate further research on the non-classical properties of multi-component quantum systems. 

\section{Acknowledgments}
The authors thank Mohamed Boubakour and Nathan Harshman for enlightening discussions. This work is supported by the Okinawa Institute of Science and Technology Graduate University (OIST). The numerical calculations were performed on the computational resources provided by the Scientific Computing and Data Analysis section at OIST. T.F., T.B., and T.D.A.T. are grateful to JST Grant No. JPMJPF2221 and T.F. also acknowledges support from JSPS KAKENHI Grant No. JP23K03290. T.D.A.T. expresses his gratitude to the Pure and Applied Mathematics University Research Institute at the Polytechnic University of Valencia for their hospitality during his visit to MAGM. MAGM acknowledges support from the Ministry for Digital Transformation and of Civil Service of the Spanish Government through the QUANTUM ENIA project call - Quantum Spain project, and by the European Union through the Recovery, Transformation and Resilience Plan - NextGenerationEU within the framework of the Digital Spain 2026 Agenda: also  from Projects of MCIN with funding from European Union NextGenerationEU (PRTR-C17.I1) and by Generalitat Valenciana, with Ref. 20220883 (PerovsQuTe) and COMCUANTICA/007 (QuanTwin), and Red Tematica RED2022-134391-T. 

\appendix

\section{Numerical Method}
\label{app:method}
In the following, we briefly summary the ab initio method used in this work, the improved Exact Diagonalization method \cite{anhtai2023quantum}. For numerical purposes, it is naturally convenient to diagonalize the many-body Hamiltonian in the formalism of second quantization. In doing so, the field operators are introduced as
\begin{align}
   \hat{\Psi}_\sigma(x) = \sum\limits_k \phi_{\sigma,k}(x)\hat{a}_{\sigma,k}, \\
   \hat{\Psi}_{\sigma}^{\dagger}(x) = \sum\limits_k \phi^*_{\sigma,k}(x)\hat{a}^{\dagger}_{\sigma,k}.
\end{align}
The field operator $\hat{\Psi}_\sigma(x)$ ($\hat{\Psi}^{\dagger}_{\sigma}(x)$) annihilates (creates) a $\sigma$-species boson being in the single-particle state $\phi_{\sigma,k}(x)$ at position $x$. Here $\hat{a}_{\sigma,k}$ ($\hat{a}^{\dagger}_{\sigma,k}$) corresponds to the annihilation (creation) operator. For bosonic systems, the creation and annihilation operators must obey the commutation relations
\begin{align}
    \left[\hat{a}_{\sigma,k},\hat{a}_{\delta,\ell}^\dagger \right] &= \delta_{\sigma\delta}\delta_{k\ell}, \\ 
    \left[\hat{a}_{\sigma,k}^\dagger,\hat{a}_{\delta,\ell}^{\dagger}\right]  &=\left[\hat{a}_{\sigma,k},\hat{a}_{\delta,\ell}\right] = 0.
\end{align}
The resulting Hamiltonian is then given as
\begin{align}
\label{full_hamiltonian_second}
\hat{H} = &\sum_{\sigma\in\{A,B,C\}}\sum_{k,\ell}  h^{\sigma}_{k\ell} \hat{a}^\dagger_{\sigma,k} \hat{a}^{}_{\sigma,\ell} \nonumber \\ &+ \dfrac{1}{2}\sum_{k\ell mn}  W^{\text{C}}_{k\ell mn} \hat{a}^\dagger_{\text{C},k} \hat{a}^\dagger_{\text{C},\ell} \hat{a}^{}_{\text{C},m}\hat{a}^{}_{\text{C},n} \nonumber \\ &+ \sum_{\sigma\neq\delta\in\{A,B,C\}}\sum_{k\ell m n}W^{\sigma\delta}_{k\ell mn} \hat{a}^\dagger_{\sigma,k} \hat{a}^\dagger_{\delta,\ell} \hat{a}^{}_{\sigma,m} \hat{a}^{}_{\delta,n}.
\end{align}
Here $h^{\sigma}_{k\ell}$, $W^\text{C}_{k\ell mn}$, and $W^{\sigma\delta}_{k\ell mn}$ are one- and two-body matrix elements  expressed in the basis of the single-particle Hamiltonian. As the system we consider is confined in a 1D harmonic trap, it is straightforward to employ the harmonic oscillator eigenfunctions as the single-particle functions $\phi_{\sigma,k}(x)$. This choice makes the uses of analytical results to obtain the matrix elements $h^{\sigma}_{k\ell} = \left(k + 0.5\right)\delta_{k\ell}$, $W^{\text{C}}_{k\ell mn} $ and $W^{\sigma\delta}_{k\ell mn}$, thus the convergence is accelerated. Further details of this approach can be found in Refs.~\cite{rotureau2013interaction,lindgren2014fermionization,dehkharghani2015quantum,anhtai2023quantum,rammelmuller2023modular}. 

The ansatz wavefunction is factorized as a linear combination of a set of orthonormal Fock states 
\begin{align}
    \label{eq:ansatze}
	|\Psi\rangle = \sum_{j_{\rm{A}}=1}^{D_\text{A}} \sum_{j_{\rm{B}}=1}^{D_\text{B}}  \sum_{j_{\rm{C}}=1}^{D_\text{C}}c_{j_{\rm{A}},j_{\rm{B}},j_{\rm{C}}} |\Phi^\text{A}_{j_{\rm{A}}}\rangle |\Phi^\text{B}_{j_{\rm{B}}}\rangle |\Phi^\text{C}_{j_{\rm{C}}}\rangle. 
\end{align}
Here, $c_{j_{\rm{A}},j_{\rm{B}},j_{\rm{C}}}$ denote the expansion coefficients and $|\Phi^{\sigma}_{j_\sigma}\rangle = |n^\sigma_1, n^\sigma_2 \dots n^\sigma_k \dots\rangle$ is a possible $\sigma$-species permanent (also known as configuration) in the total of $D_\sigma$ permanents used to expand the ansatz. Note that in each permanent, the occupation number $n^\sigma_k$ in the single-particle state $\phi_{\sigma,k}(x)$ can be arbitrary integers between 0 and $N_\sigma$ and must obey the constraint $\sum\limits_k n_k^\sigma = N_\sigma$. In practice, the permanents for the ansatz are selected such that their energies in the non-interacting many-body Hamiltonian are less than a certain optimal value $E_{max}$. This truncation approach is known as the energy-cutoff scheme proposed in Refs.~\cite{chrostowski2019efficient,plodzien2018numerically} and hence the accuracy of the numerical results is controlled by increasing $E_{max}$. The key ingredient of the improved ED scheme which significantly reduces the number of required permanents is the selection of dominant configurations in terms of the spatial symmetry of the desired many-body state \cite{anhtai2023quantum}. It is known that if the trapping potential of a quantum particle has a spatial symmetry, namely
\begin{equation}
    V(x) = V(-x),
\end{equation}
then the single-particle eigenfunctions $\phi_k(x)$ possess a well-defined spatial symmetry. Mathematically, the spatially symmetric $\phi_{k=2n}(x)$ are even functions, while the spatially asymmetric $\phi_{k=2n+1}(x)$ are odd functions. Because the permanents are the symmetrized Hartree product of the single-particle functions $\phi_k(x)$, they must satisfy this spatial symmetry. This leads to the fact that the desired wave functions are solely expanded by either even- or odd-parity permanents. Thus, it is more practical to use only the permanents that have the same parity as the desired many-body wavefunction in the expansion of the ansatz. Since the bosonic ground-state wavefunction is spatially symmetric, we therefore only use even-parity permanents for the ansatz in this work. Meanwhile, for obtaining the first excited state, the ansatz is expanded by odd-parity permanents. 

In general, minimizing the expectation value of the Hamiltonian \eqref{full_hamiltonian_second} with respect to the ansatz \eqref{eq:ansatze} leads to standard Hermitian eigenvalue problem which can be written in the form of
\begin{equation}
	\label{manybody_equations}
	\mathbf{H}|C_m\rangle = E_m|C_m\rangle,
\end{equation}
with $\mathbf{H}$ being the matrix representation of the many-body Hamiltonian \eqref{full_hamiltonian_second}. The pair $\{E_m, |C_m\rangle \}$ is the $m$-th many-body eigenvalue and eigenvector, respectively. 

\bibliography{Biblio.bib}
\end{document}